\documentclass[aps,prd,12pt,superscriptaddress,
amsfonts,amssymb,amsmath,byrevtex,showpacs,nofootinbib]{revtex4-1}
\usepackage{graphicx}
\usepackage{amssymb,amsmath,amsfonts,palatino,amsthm}
\usepackage{amssymb}
\usepackage{epstopdf}
\usepackage{color}

\begin{document}

\title{Loop-deformed Poincar\'e algebra\footnote{Essay written for the Gravity Research 
Foundation 2013 Awards for Essays on Gravitation. \\ Submitted on 30 March 2013. \\ }}

\author{Jakub Mielczarek \\
Institute of Physics, Jagiellonian University, Reymonta 4, 30-059 Cracow, Poland \\
Department of Fundamental Research, National Centre for
Nuclear Research,\\ Ho{\.z}a 69, 00-681 Warsaw, Poland}

\email{jakub.mielczarek@uj.edu.pl}

\begin{abstract}

In this essay we present evidence suggesting that loop quantum gravity
leads to deformation of the local Poincar\'e algebra within the limit of high energies. 
This deformation is a consequence of quantum modification of effective 
off-shell hypersurface deformation algebra. Surprisingly, the form of deformation 
suggests that the signature of space-time changes from Lorentzian to Euclidean 
at large curvatures. We construct particular realization of the loop-deformed Poincar\'e
algebra and find that it can be related to curved momentum space, which indicates 
the relationship with recently introduced notion of relative locality. The presented 
findings open a new way of testing loop quantum gravity effects. 

\end{abstract}

\maketitle 

In crystals, symmetries of continuous rotations and translations are broken due to 
discrete nature of the molecular lattice structure. Similarly, it is usually expected 
that isometries of Minkowski space, described by the Poincar\'e group, are broken 
or deformed due to granularity of space at the Planck scale. Nevertheless, one 
can imagine that some types of discontinuity may preserve the large scale symmetries. 

Let us see what it looks like in loop quantum gravity (LQG) \cite{Ashtekar:2004eh}, 
which is a background independent approach to quantize gravity. In LQG, ``atomic'' nature 
of space is reflected by discrete spectra of geometrical operators, such as area and volume. 
However, this feature alone does not guarantee violation of Lorentzian symmetry. Similarly,  
a discrete spectrum of square of the angular momenta operator does not imply breaking of 
the rotational invariance. Indeed, as pointed out by Rovelli and Speziale \cite{Rovelli:2010ed}, 
quantum amplitudes of LQG are manifestly Lorentz covariant. Their result concerns kinematical 
sector of the theory comparably to discreteness of spectra of the geometrical operators.  
However, LQG is not only kinematics but also constraints that must be imposed in order to extract 
physical states. Furthermore, the off-shell algebra of these constraints carries information about 
symmetries of the theory.  

At the classical level, this algebra corresponds to hypersurface deformation algebra, encoding
general covariance,  a cornerstone of general relativity (GR).  However, the fate of this algebra at the 
quantum level is unknown due to the complicated form of the quantum constraints. 
Nevertheless, significant progress has recently been made in analysis of the effective off-shell 
algebra, where LQG effects were introduced through systematic modifications of the classical 
constraints. This analysis was performed for both, spherically symmetric configurations 
\cite{Reyes,Bojowald:2011aa} and perturbative inhomogeneities on the flat Friedmann-Robertson-Walker 
(FRW) background \cite{Bojowald:2008gz,Cailleteau:2011kr}.  Main message coming from these 
studies is that the algebra of effective constraints is deformed with respect to its classical counterpart. 
Namely, it takes the following form:  
\begin{eqnarray}
\left\{D[M^a],D [N^a]\right\} &=& D[M^b\partial_b N^a-N^b\partial_b M^a], \nonumber \\
\left\{D[M^a],S^Q[N]\right\} &=& S^Q[M^a\partial_b N-N\partial_a M^a],  \nonumber  \\
\left\{S^Q[M],S^Q[N]\right\} &=& {\color{blue} \beta} 
D\left[q^{ab}(M\partial_bN-N\partial_bM)\right], \nonumber
\end{eqnarray}
where $\beta$ is a deformation factor, equal one in the classical theory with Lorentzian signature. 
Superscript $Q$ indicates that scalar constraint $S$ is quantum corrected, whereas diffeomorphism 
constraint $D$, in the spirit of LQG, holds the classical form. 

Functional form of the factor $\beta$ depends on whether the so-called \emph{inverse volume} 
or \emph{holonomy} quantum corrections are applied.  For inverse volume corrections, 
$\beta$ depends on spatial metric $q_{ab}$, while holonomy corrections introduce dependence 
on extrinsic curvature. Interestingly, for both types of corrections, $\beta$ falls to zero within the regime 
of strong quantum gravitational effects. In consequence, the algebra of constraints becomes reduced 
to the \emph{ultralocal} form \cite{Isham:1975ur} which prohibits any causal contacts  between space points.  
Because information (waves) cannot propagate spatially, this phase of gravity is also known as 
\emph{asymptotic silence} \cite{Mielczarek:2012tn}. What is extremely interesting, is that such fancy 
behavior is expected also in the high curvature limit of GR. This phenomenon is known as 
Belinsky-Khalatnikov-Lifshitz (BKL) conjecture \cite{BKL,Andersson:2004wp}. 
 
However, that is not all. In case of holonomy corrections, one can cross the ultralocal stage and go to 
regime of negative $\beta$. Before we interpret what it means let us firstly say a few words on 
the origin of holonomy corrections. To be more specific, let us focus on example of FRW cosmology. 
In this case, holonomy corrections are introduced  formally by replacing extrinsic curvature $\bar{k}$ by 
$\sin(\delta \bar{k})/\delta$ function, with $\delta$ being a function of a scale factor. Because the 
$\delta \bar{k} \rightarrow \delta \bar{k} +2\pi$ symmetry is introduced, the procedure can be seen as 
periodification of the real line to $U(1)$. The classical limit is recovered for $\delta \bar{k} \ll 1$

It was shown that for both, spherically symmetric models and cosmological perturbations 
with holonomy corrections, the deformation factor $\beta$ takes the same \emph{cosine} form. 
Additionally, in case of cosmological perturbations, we can express $\beta$ in terms of matter 
energy density $\rho$  \cite{Cailleteau:2011kr}:   
\begin{equation}
\beta=\cos(2\bar{\mu} \gamma\bar{k}) = 1 - 2\frac{\rho}{\rho_{\text{c}}} \in [-1,1], \nonumber
\end{equation}
where maximal energy density $\rho_{\text{c}} = \frac{3}{8\pi G \lambda^2 \gamma^2} \sim \rho_{\text{Pl}}$ and
the Planck energy density $\rho_{\text{Pl}}\equiv E_{\text{Pl}}^4$, with $E_{\text{Pl}} \approx 1.22 \cdot 10^{19}$ GeV.
It is clear that at low energy densities $(\rho \ll \rho_{\text{c}})$ the classical limit ($\beta=1$) is recovered. 
However, while passing to regime of high energy densities, deviations from the classical theory are 
evident. Firstly, while approaching $\rho =  \rho_{\text{c}}/2$ we ``slow down" to the state of asymptotic silence. 
Secondly, for $\rho < \rho_{\text{c}}/2$, sign of $\beta$ reverses, which is associated with change of the metric 
signature from Lorentzian to Euclidean \cite{Mielczarek:2012pf}. Amazingly, the possibility of Euclidezation of space-time
due to quantum gravitational effects was anticipated already by Hartle and Hawking in their famous no-boundary 
proposal \cite{Hartle:1983ai}.    
  
What we have learned so far is that the effective algebra of constraints is deformed due to loop quantum 
gravity effects in a very interesting way. Consequently, general covariance is deformed and because Lorentz 
transformations and translations are special types of coordinate transformations, we expect that algebra 
of generators of these transformations is deformed as well. The crucial observation made by Bojowald and 
Paily was that this algebra can be recovered from the hypersurface deformation algebra by considering 
linear laps function $N$ and the shift vector $N^a$ \cite{Bojowald:2012ux}:
\begin{equation}
N(x)= \Delta t+v_a x^a,  \ \ N^a(x)=\Delta x^a +{R^a}_b x^b, \nonumber
\end{equation}
together with the flat spatial metric $q_{ab}=\delta_{ab}$. Taking this into account, they showed that the resulting 
deformation of the Poincar\'e algebra is a special case of a broader class of deformations, with modifications of 
generators of boosts, studied in Ref. \cite{Kovacevic:2012an}. However, as we show here, modification of boosts is 
not required if $\beta$ depends on generator of time translation only.  In that case,  the loop-deformed Poincar\'e 
algebra can be obtained from the following Heisenberg algebra 
\begin{equation}
[X_{\mu},X_{\nu}] = 0 ,\ \ \ [P_{\mu},P_{\nu}] =0, \ \ \ [X_{\mu},P_{\nu}] = i g_{\mu\nu}(P_0), \nonumber
\end{equation}
where $g_{\mu\nu}(P_0) = \text{diag}(-\beta(P_0),1,1,1)$. By introducing classical generators of rotations 
$J_i \equiv \frac{1}{2} \epsilon_{ijk} (X_jP_k-X_kP_j)$ and boosts $K_i \equiv  X_iP_0-X_0P_i$, we find that the 
following algebra is fulfilled: 
\begin{eqnarray}
\left[J_i,J_j \right] &=& i \epsilon_{ijk} J_k, \ \ \ 
\left[J_i,K_j \right] =  i \epsilon_{ijk} K_k, \ \ \ 
\left[K_i,K_j \right] =  -i {\color{blue} \beta(P_0)} \epsilon_{ijk} J_k, \nonumber \\
\left[J_i,P_j \right] &=& i  \epsilon_{ijk} P_k, \ \ \ 
\left[K_i,P_j \right] =  i \delta_{ij} P_0, \ \ \ 
\left[J_i,P_0 \right] = 0, \nonumber  \\
\left[K_i,P_0 \right] &=& i {\color{blue} \beta(P_0)}P_i,\ \ \ 
\left[P_i,P_j \right] = 0, \ \ \ 
\left[P_i,P_0 \right] =  0. \nonumber 
\end{eqnarray}
This algebra has the same form as the one found in Ref. \cite{Bojowald:2012ux}. However, here the boosts 
are classical and $\beta$ depends on $P_0$ only.  The classical  Poincar\'e algebra is recovered for 
$\beta \rightarrow 1$, while for $\beta=0$ we obtain Carrollian limit \cite{Levy} describing ultralocal 
space-time. Furthermore, for $\beta=-1$, the algebra describes isometries of the 4D Euclidean 
space. Therefore, we have full compatibility with the conclusions derived at the level of algebra of constraints. 

It is worth stressing that, while the loop-deformed Poincar\'e algebra is recovered from the algebra of 
constraints in the limit of flat space-time, the metric entering Heisenberg algebra reveals explicit 
dependence on $P_0$. This can be associated with curvature of momentum space, which suggests 
relation of our approach with recently introduced concept of \emph{relative locality} 
\cite{AmelinoCamelia:2011bm,AmelinoCamelia:2011pe}.  An unavoidable consequence of the momentum 
space curvature is modification of dispersion relation of particles, \emph{e.g.} photons.  
      
Because the first Casimir invariant for our loop-deformed Poincar\'e is 
\begin{equation}
\mathcal{C}_1 =-\int^{P_0}_0 \frac{2y}{\beta(y)}dy+{P_i}^2,  \nonumber
\label{Casimir1def}
\end{equation}
the dispersion relation of photon takes the form $\int^{E}_0 \frac{2y}{\beta(y)}dy={\bf p}^2$, where we 
defined $E=P_0$ and  ${\bf p}^2=(P_i)^2$. For $\beta=1$, the classical relation $E^2={\bf p}^2$ is 
correctly recovered. By differentiating the deformed dispersion relation with respect to $E$ one can 
show that $v_{\text{gr}}v_{\text{ph}}=\beta$, where $v_{\text{gr}}$ and $v_{\text{ph}}$ are 
group and phase velocities respectively. These velocities are no longer constant but vary as a function 
of energy.  Furthermore, under the following assumptions: $0<\beta \leq 1$, $\beta(0)=1$ and 
$\frac{d\beta}{dE} \leq 1$, one can prove that they never exceed the speed of flight in 
vacuum ($v_{\text{ph}} \leq 1$, $v_{\text{gr}} \leq 1$).

\begin{figure}[ht!]
\centering
\includegraphics[width=11cm,angle=0]{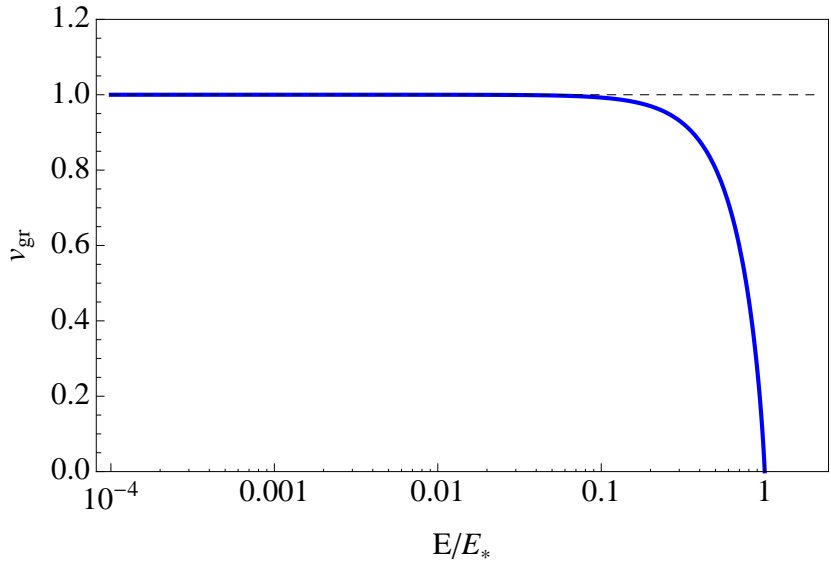}
\caption{Group velocity of photons is represented here as a function of energy for the model with
deformation function $\beta(E) =1-\frac{E^2}{E^2_*}$. The group velocity 
falls to zero while $E\rightarrow E_*$, which corresponds to the Carrollian limit ($\beta \rightarrow 0$).} 
\label{GrVel}
\end{figure}

As an example, let us take $\beta(E) =1-\frac{E^2}{E^2_*}$. However, for other choices, \emph{e.g.} $\beta(E) =\cos(E/E_*)$, 
results are qualitatively the same. The corresponding energy dependence of the group velocity, 
given by formula 
\begin{equation}
v_{\text{gr}} = \left(1-\frac{E^2}{E^2_*}\right)\frac{E_*}{E} \sqrt{-\log\left(1-\frac{E^2}{E^2_*}\right)}
= 1-\frac{3}{4}  \left(\frac{E}{E_*}\right)^2+\mathcal{O}(E^4), \nonumber
\end{equation}
is shown in Fig. \ref{GrVel}. 
The free parameter $E_*$ can be constrained  by studying time lags of high energy photons arriving 
from the distant astrophysical sources, such as gamma ray bursts (GRB)  \cite{AmelinoCamelia:1997gz}. 
In our case, this time lag is expressed as $\tau \simeq \frac{L}{c}\frac{3}{4}  \left(\frac{E}{E_*}\right)^2$,  
where $L$ is a distance to source.  Using constraint from the GRB 090510 detected by the 
Fermi satellite \cite{Ackermann:2009aa} we find that $E_*> 4.7 \cdot 10^{10}$ GeV.  The obtained constraint
is weak relative to the Planck scale because quantum corrections are quadratic here, as suggested by the cosine 
form of $\beta$. Nevertheless, a new possibility of testing loop quantum gravity emerges. What is especially worth 
stressing is that both, dynamics of very early universe and propagation of astrophysical particles are affected by 
the same deformation $\beta$. This provides a unique possibility of constraining the same physics by using 
completely different observations. This might open a new stage in cosmic search for the quantum gravity effects.

\end{document}